\documentclass[twocolumn,aps,pre,showpacs,amsmath,amsfonts,amssymb,floatfix]{revtex4}
\usepackage{morefloats}
\usepackage{amsmath}
\usepackage{graphicx}
\usepackage{dcolumn}
\usepackage{bm}

\begin{document}

\title{Emergence of clustering: Role of inhibition}

\author{Sanjiv K. Dwivedi and Sarika Jalan\footnote{sarika@iiti.ac.in}}
\affiliation{Complex Systems Lab, School of Basic Sciences, Indian Institute of Technology Indore,
M-Block, IET-DAVV Campus Khandwa Road, Indore-452017 }

\begin{abstract}
Though biological and artificial complex systems having inhibitory connections
exhibit high degree of clustering in their interaction pattern, the evolutionary origin of clustering in such systems remains 
a challenging problem. 
Using genetic algorithm we demonstrate that inhibition is required in 
the evolution of clique structure from primary random architecture, in which 
the fitness function is assigned based on the largest eigenvalue. 
Further, the distribution of triads over nodes of the network evolved from mixed connections reveals a negative 
correlation with its degree providing insight into origin of
this trend observed in real networks.
\end{abstract}
\maketitle 

Structural features of interaction patterns in complex systems are not completely random, they possess some non-random part, possibly dynamical response dependent local or global structures\cite{community}. 
Several models as well as statistical measures have been proposed to 
quantify specific features of networks like degree distribution, small world property, 
community structure, assortative or disassortative mixing etc. 
Abundance of cliques of order three, indicated by high clustering
coefficient ($C$), plays a crucial role in organizing local motif structures that enhance the robustness of the underlying system 
\cite{alon_book}. The functional roles of such motifs have been
intensely studied \cite{Alon}. Many biological networks such as 
metabolic \cite{Ravasz}, transcription \cite{Potapov}, protein-protein 
interaction \cite{Yeger}, neuronal systems \cite{Strogatz},
food-web \cite{rev_barabasi} and social systems \cite{Newman,SJ_PLoS} are rich in
the clique structure. Moreover,
the local $C$ of nodes have been found to be negatively correlated with
their degree in metabolic networks \cite{Ravasz}.
This paper presents a novel method
to understand the evolution of clustering
and distribution pattern of cliques over nodes which are known
to lead hierarchical 
organization of modularity in network. 
Here we use stability criteria for genetic algorithm to choose
from the population which leads to the evolution of clustering in the
final network. We find that presence of inhibitory links during evolution
is very crucial for evolution of clustering.

Previous attempts to provide
evolutionary understanding of
emergence of cooperation \cite{Ferreira} as well as
to use clustering
based constraints for the evolution of other structural properties \cite{Alexi} 
fail to
incorporate effect of inhibition in the connection. 
Coexistence of inhibitory and excitatory couplings have been implicated in various systems.
For instance, in ecosystems, competitive, predator-prey and mutualistic interactions exist among communities of species \cite{MayNature1972}. Excitatory (friendly) and inhibitory (antagonistic) interactions are also evident in social systems \cite{Leskovec}. In neural networks, excitatory and inhibitory synapses regulate the potential variations in neural populations \cite{book_brain}.
In context of ecological systems, a celebrated work by Robert May demonstrates that
largest real part of eigenvalues ($R_{\mathrm {max}}$)
of corresponding adjacency matrix, determined by equal contribution from connectivity and
disorder in coupling strength
contains information about stability of the underlying system \cite{MayNature1972}. Spectral properties for matrices of ecological and metabolic systems 
have been further
shown to be useful for determining
stability criteria based on their interaction properties \cite{Allesina,Metabolic}.
This notion has further been propagated for neural
networks where eigenvalues with larger real part destabilize the silent state 
of the system \cite{Rajan}. 
Recent work has demonstrated that the fluctuations of $R_{\mathrm {max}}$ leads to transition to the extreme value statistics at particular ratio of inhibitory couplings further emphasizing the  importance of inhibitory connections
in networks \cite{Extreme}. 

Genetic algorithm (GA) is a randomized technique motivated from the natural selection process encountered in a species in course of its evolution, that has been successfully applied to computational problems dealing with exponentially large search space \cite{Holland}
as well to model evolutionary systems \cite{GA_forrest}. 
Evolution of hierarchical modularity in random directed networks has been proposed \cite{Variano}. However this approach has been reported to be seemingly insufficient to produce modularity \cite{Kashtan},
hence the introduction of clustering leading to formation of local structures, might help to refine our understanding pertaining to evolution of hierarchical organization.
Instead of directed coupling and segregation of system during generations \cite{Variano},
we consider bidirectional coupling in connected systems with average degree and connectivity being conserved during evolution and investigate the role of inhibitory and excitatory connections behind the 
existence of clustering in interaction patterns in the evolved network.

Motivated by the coupling behavior known for many real world systems
that for a given pair of
individuals the behavior remains fixed, we randomly assign behavior to individual nodes of each pair in a time 
invariant fashion. For instance in food web, the nature of interaction between any pair 
of predator-prey remains fixed \cite{Schaffer}. Also in the 
mutualistic association of fungi and roots of vascular plant 
ecosystem, the parasite benefits at the expense of the host \cite{Barbosa}.
Furthermore, we introduce randomness in connection strength which fluctuates with respect to 
time \cite{Allesina,Yuan}. For assigning random weights we choose a uniform random variable, 
however our proposed technique stands valid for other
random variables as well. We implement the above assumptions in GA, elaborated as under.  

Considering the Erd\"os-R\'enyi (ER) random undirected sparse networks \cite{rev_barabasi} as the initial population, we generate another matrix ([$b_{\mathrm {ij}}$]), devoid of zero entries,
consisting of randomly assigned `+1' and `-1' entries (`-1' entries being assigned with probability $p_{\mathrm {in}}$) in order to define behavior of links during evolution. 
If a link is assigned positive or negative value, it will carry the same sign throughout,
and evolution affects only strength of the connection.
The fitness of a network belonging to population used in GA, is defined on the basis of $R_{\mathrm {max}}$ of matrix ([$c_{\mathrm {ij}}$]), constructed using its sparse adjacency matrix ([$a_{\mathrm {ij}}$])  and matrix [$b_{\mathrm {ij}}$]. 
Note that the largest modulus of the eigenvalues 
$\lambda_{max}$ for adjacency matrices of undirected networks characterizes various dynamical properties like threshold of 
phase transition in virus spread \cite{largest_virus}
as well as synchronization of coupled 
oscillators \cite{largest_syn}. 
The GA in this paper minimizes $R_{max}$ as it quantifies the stability of
underlying system. In case
of symmetric matrices having all real eigenvalues, as well as 
for the matrices with non-negative entries, according
to the Perron Frobenius theorem $R_{max}=\lambda_{max}$. 
However, for asymmetric matrices with positive and negative entries, both the quantities $R_{max}$ and $\lambda_{max}$ are distinct. 
In case of predator-prey interactions, due to elliptical shape of spectra, 
the major axis lies on the imaginary axis, and despite higher value of $\lambda_{max}$ 
due to the larger imaginary part of eigenvalues,
stability of the system has been shown to be characterized by $R_{max}$
\cite{Allesina}. 
Furthermore, construction of matrix [$c_{\mathrm {ij}}$] is inspired by random behavior
of coupling strength in real world networks.
\begin{equation}
c_{\mathrm {ij}} = \begin{cases} b_{\mathrm {ij}}X~~\mbox{if } a_{\mathrm {ij}}\not=0 \\
0 ~~ \mbox{otherwise} ~~a_{\mathrm {ij}}=0. \end{cases}
\label{rec_mat}
\end{equation}
where $X$ is a uniform random number between 0 and 1.

We arrange the networks on the basis of increasing order of 
$R_{\mathrm {max}}$ values of their associated matrices([$c_{\mathrm {ij}}$]).
For the next time step, the top 50$\%$ of the networks having lower $R_{\mathrm {max}}$ values, termed as fitter networks, are filtered.  
In the next time step, these fitter networks are considered
and the remaining 50$\%$ of the networks are constructed
by generating cross of randomly selected pairs of fitter networks. Such a cross is created by randomly selecting blocks of adjacency matrices of specified dimension (10 in this case) with equal probability. Undirected networks are constructed by considering the upper triangular part of 
these crossed matrices. The average degree of the associated crossed child network is maintained   
by randomly removing or inserting connections in the networks with some probability (decided by fluctuation in the crossed population with expected total degree of the initial random network). Next we check for fluctuations in absolute value of differences between the mean of $C$ of the pairs of randomly selected fitter networks and the $C$ of their crossed child networks. Only small fluctuations are taken in to consideration. 
On encountering large fluctuation, we discard the generated child network and repeat the preceding steps for creating a cross with the same pair of
the fitter networks. Small fluctuations are considered, so that the child networks conserve the property ($C$) inherited from their parents \cite{Montana}.
The above mentioned procedure is repeated for the desired number of time steps. 
\begin{figure}[t]
\centerline{\includegraphics[width=\columnwidth]{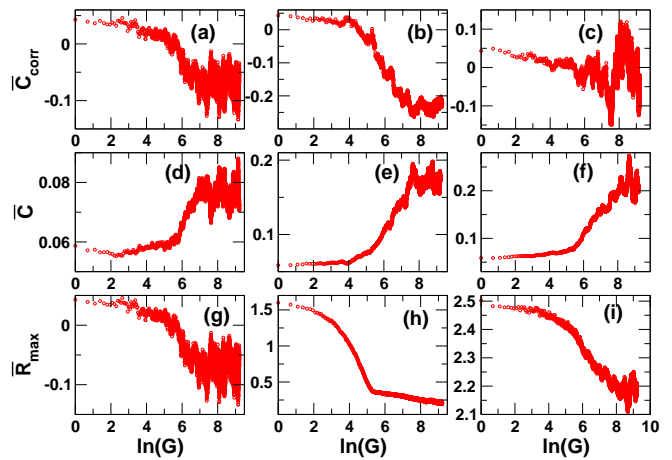}}
\caption{(Color online) (a), (d) and (g) plot evolution of average $C$ average $C_{\mathrm {corr}}$ 
and average of $R_{\mathrm {max}}$ respectively over generations (G) for system having only excitatory connections($p_{\mathrm {in}}$ = 0). The average is taken over population used in GA. Similarly (b), (e) and (h) depict evolution for system having mixed of the both types
(excitatory and inhibitory) of connections($p_{\mathrm {in}}$ = 0.5). 
(c), (f) and (i) show evolution for system having only inhibitory type of connections($p_{\mathrm {in}}$ = 1.0).
For all the cases, initially taken ER networks have average degree 6 
with $N=100$.}
\label{Fig1}
\end{figure}

In a system with only excitatory connections, i.e $p_{\mathrm {in}}$ = 0, weak cluster formation is observed (Fig.~\ref{Fig1} (d)) with
respect to the minimization of $\overline{R}_{\mathrm {max}}$ values during the evolution (Fig.~\ref{Fig1} (g)). This slight increase in $\overline{C}$ (Fig.~\ref{Fig1} (d)) is due to the random fluctuation of coupling weight. Devoid of such random fluctuations, cluster formation is not found. 
We note that for only excitatory connections the decrease in $\overline{R}_{\mathrm {max}}$ is also not significant as evolution progresses. The correlation of local clustering coefficient of a node with its degree($\overline{C}_{\mathrm {corr}}$) exhibits
a convergence towards weak negative values. The decrease or increase in the values of 
($\overline{C}_{\mathrm {corr}}$, $\overline{R}_{\mathrm {max}}$, $\overline{C}$) exhibits a smooth variation during the structural evolution up to a certain saturation value, after which random fluctuation is observed due to the
random variation in the coupling strength during the evolution.
\begin{figure}[t]
\centerline{\includegraphics[width=\columnwidth]{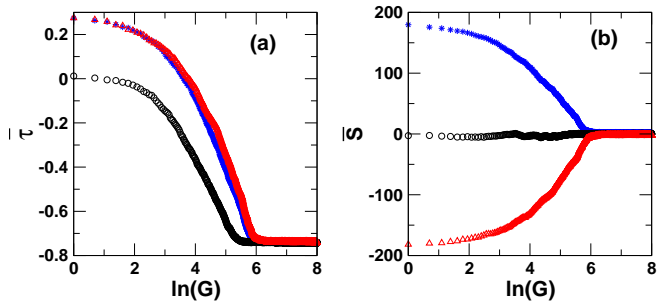}}
\caption{(Color online) Evolution of the average value of $\tau$ and sum of entries of the matrices([$c_{\mathrm {ij}}$])
during evolution for $p_{\mathrm {in}}$ = 0.20, $p_{\mathrm {in}}$ = 0.50 and $p_{\mathrm {in}}$ = 0.80
depicted with triangle (red), circle (black) and star (blue) respectively.
Initial ER networks have average degree 6 with $N=100$.}
\label{Fig2}
\end{figure}

\begin{figure}[h]
\centerline{\includegraphics[height=4.0cm,width=6cm]{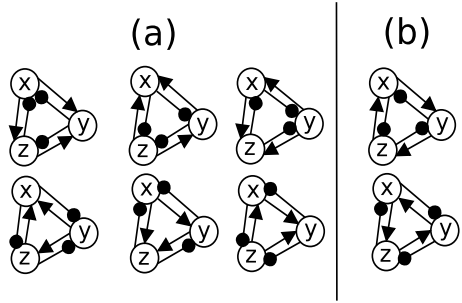}}
\caption{(Color online) Two types of triads for mixed case with $p_{\mathrm{in}}$ = 0.50 which include coupling behavior evolved in [$c_{\mathrm {ij}}$] matrices. Dots and arrows represent
inhibitory and excitatory links, respectively. }
\label{Fig3}
\end{figure}

In a system comprising of both excitatory and inhibitory connections (mixed case), for $p_{\mathrm {in}}$ = 0.5, prominent clustering is seen even when average degree remains fixed (Fig.~\ref{Fig1} (e)).
As the value of $\tau$ ($\tau$ = $\sum_{i,j=1}^{N} c_{\mathrm{ij}}c_{\mathrm{ji}}/\sum_{i,j=1}^{N} c_{\mathrm{ij}}c_{\mathrm{ij}}$) decreases, the $R_{\mathrm {max}}$ value also decreases, since the spectral distribution of associated matrices are elliptical in shape and its axis on real line decreases around a fixed center \cite{Sompolinsky}.
The rate of decrement in  
$\overline{R}_{\mathrm {max}}$ values is high (Fig.~\ref{Fig1} (h)) till certain generations due to the steep decrease in the values of $\tau$, after which 
there is a saturation
(Fig.~\ref{Fig2}(a)) .
Additionally, average of $\tau$ taking negative values (Fig.\ref{Fig2}(a))
reflects an increase in
the antisymmetric (predator-prey) type of couplings. Such types of couplings are known to play an instrumental 
role in conferring robustness to ecological systems under
external perturbations \cite{Stefano}.
We find that the triads, which significantly contribute to the clustering of a network,
are evolved in the [$c_{\mathrm{ij}}$] matrices depending upon if excitation (inhibition) from a particular node
in a triad gets carried forward and comes back to the same node (Fig.\ref{Fig3}(b)) or not (Fig.~\ref{Fig3}(a)), it turns out that type (a) constitute $\sim75\%$ of the triads in the evolved network, 
while the rest are of type (b). As discussed above, in the 
course of evolution of [$c_{\mathrm{ij}}$] matrices through the GA algorithm, the dominant behavior that persists towards the end is the predator-prey type interactions and the triads observed are of (a) and (b) types. All other types of interactions, for example, competitive (inhibitory-inhibitory) and cooperative (excitatory-excitatory), which might be present in the initial networks, evolve to form triads of type (a) and (b) eventually, where type (a) hails six possible conformations and type (b) adopts only two conformations leading to equivalent probability of occurrence of these triads in the evolved networks.
In real-world ecological systems, triads of type (a), popularly known as omnivory chain, are found in abundance \cite{Stouffer,Jordi}
and the explanation behind its origin might be an interesting problem to investigate. 

Further, the $\overline{C}_{\mathrm {corr}}$ decreases consistently, depicting the realistic
clique distribution over the nodes. This evolved feature of simulation-driven networks complies with the interesting power law behavior followed by the degree and associated clustering coefficient in biological systems \cite{Ravasz}.

Moreover, the slower rate of decrease of  $\overline{R}_{\mathrm {max}}$ does not affect the rate of structural changes captured in terms of clustering (Fig.~\ref{Fig1} (e)). The value of $\overline{C}$ 
increases with constant rate and after attaining a fixed configuration, fluctuates due to the random fluctuations in the coupling strengths.
It is observed that the mean values of  $R_{\mathrm {max}}$ lies very close to its minima over the population. However, the maxima of
 $R_{\mathrm {max}}$ is widely separated from the minima and mean.
The minimum, mean and maximum values of $C$ over the population of the networks used in GA increase together.
Surprisingly, the systems consisting of only inhibitory couplings ($p_{\mathrm {in}}$=1) display the
higher values of $\overline{C}$ over the evolution
as compared to the
systems dealing with only excitatory connections (Fig.~\ref{Fig1} (f)).
The rate of decrement of  $\overline{R}_{\mathrm {max}}$ values is smooth as compared to the mixed case.  
However the maxima, mean and minima of  $\overline{R}_{\mathrm {max}}$ are consistently separated during the evolution. In this case also, the minimum, mean 
and maximum values of $C$ over the population of the networks used in GA increase together. As opposed to the system with excitatory connections, even after removal of the constraints of random fluctuations in coupling strength, the system in this case exhibits clustering. Recently emergence of modularity has been studied in competitive
interactions using dynamical aspects of Kuramoto model \cite{Salvatore}. The fact that modular structures have high $C$ values, supports our finding. Furthermore, the evolved system does not show  $\overline{C}_{\mathrm {corr}}$ convergence to fixed values and its average value always fluctuates about zero. The anomalous behavior of  $\overline{C}_{\mathrm {corr}}$ may be further explored to have evolutionary understanding of real world systems.

Even with an increased inhibitory or excitatory connections in the matrix [$b_{\mathrm {ij}}$], the system proceeds towards the balanced situation where inhibitory connections are counterbalanced by the excitatory connections. 
Fig.~\ref{Fig2}(b) depicts two different paths for higher excitatory (star) and higher
inhibitory (triangle) couplings to unite at balanced situation. For $p_{\mathrm {in}}$ = 0.50, the balanced situation is
maintained over generations. The measure  $\overline{\tau}$ decreases for all the three cases and converges to a single value (Fig.~\ref{Fig2}(a)). 
\begin{figure}[t]
\centerline{\includegraphics[width=0.8\columnwidth]{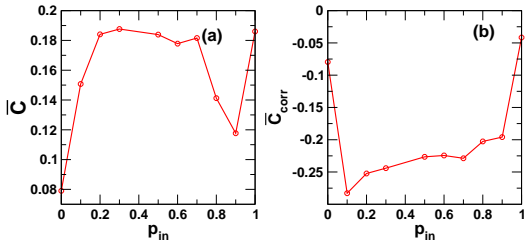}}
\caption{(Color online) Evolved average value of $C$ and $C_{\mathrm {corr}}$ for different $p_{\mathrm {in}}$ values in panels (a) and (b)
respectively. In this case system is evolved till 5000 iterations and average of $C$ and $C_{\mathrm {corr}}$ is taken over population and again average is
taken over last 1000 generations. Initial ER networks have
 $\langle k\rangle$=6 and $N=100$.}
\label{Fig4}
\end{figure}
As $p_{\mathrm {in}}$ increases the $\overline{C}$ values increase at a faster pace up to $p_{\mathrm {in}}$ = 0.2, after
which they saturate(Fig.~\ref{Fig4}(a)). 
This value is approximatively maintained till $p_{\mathrm {in}}$ = 0.7 after
which $\overline{C}$ value decreases until it reaches to its minimum. 

\begin{figure}[h]
\centerline{\includegraphics[width=0.9\columnwidth]{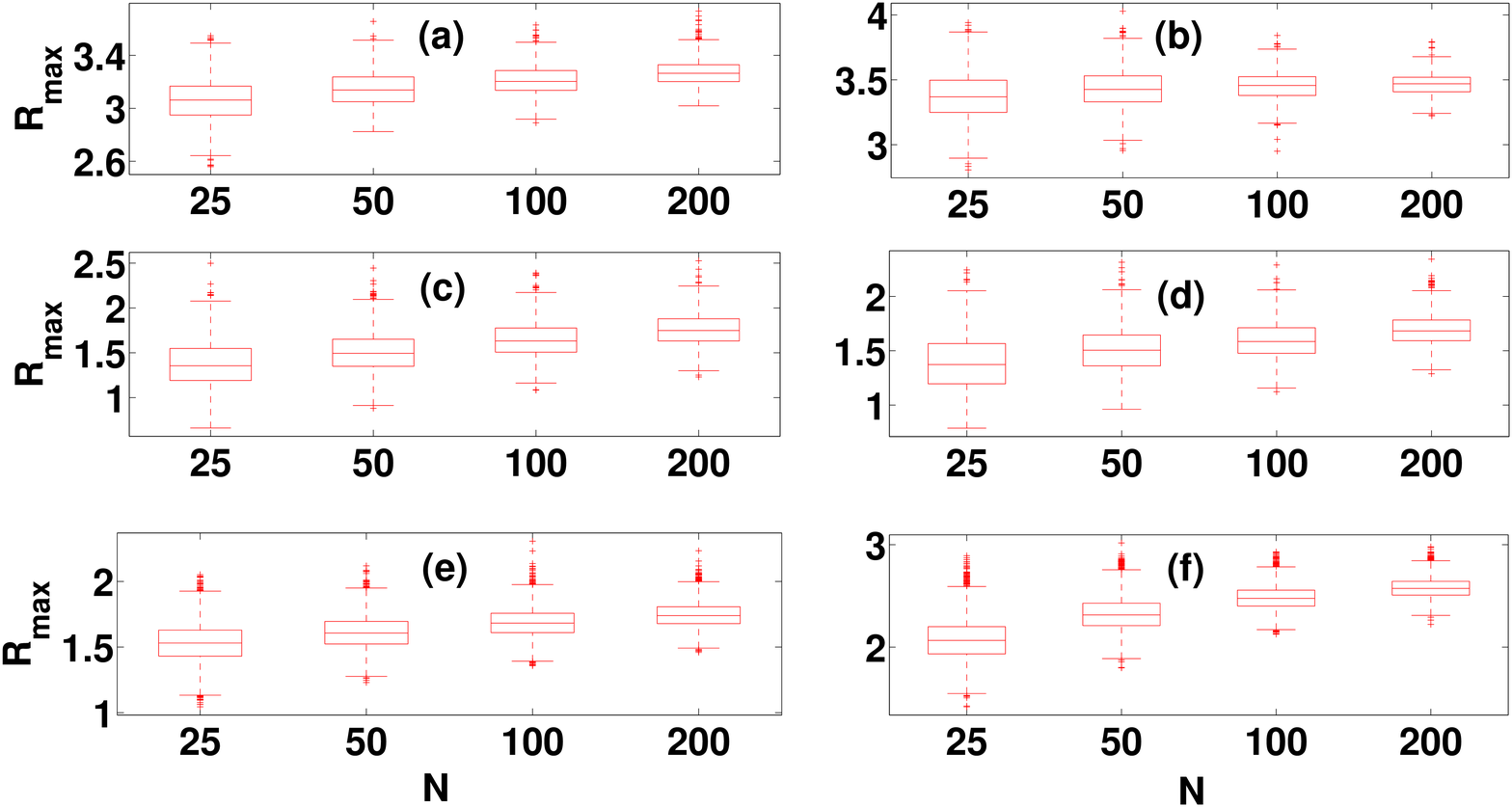}}
\caption{(Color online) (a) and (b) $R_{\mathrm {max}}$ values for lattices and random networks respectively with system having only excitatory types of connections. Similarly (c) and (d) are for mixed types of connections, and
(e) and (f) for only inhibitory type of connections.}
\label{Fig5}
\end{figure}

$\overline{C}_{\mathrm {corr}}$ values attain the lowest point at  
$p_{\mathrm {in}}$ = 0.10, after which they increase at a slower pace over a 
long regime  of $p_{\mathrm {in}}$ followed by the increase at a higher pace
for $p_{\mathrm {in}}$ $>$ 0.7, finally exhibiting a sharp increase 
$p_{\mathrm {in}}$ = 0.9 onwards (Fig.\ref{Fig4}(b)). 
The plausible explanation of the acute decrease of $\overline{C}$ values 
from $p_{\mathrm {in}}$ = 0.70 in Fig.\ref{Fig4}(a) is as follows.
Despite very high $p_{\mathrm {in}}$ value, as evolution progresses, 
the networks displays the balance between inhibitory and excitatory couplings with 
a faster rate than the evolution of  $\overline{C}$ and
 $\overline{C}_{\mathrm {corr}}$ (Fig.\ref{Fig2}(b)). 
The values of  $\overline{C}_{\mathrm {corr}}$ converges towards zero 
(for $p_{\mathrm {in}}$ $>$ 0.7).  Note that for lattices, 
$C_{\mathrm {corr}}$ value is zero and overlapping region of 
$R_{\mathrm {max}}$ spread for a network having
a balance between inhibitory and excitatory coupling
is much higher than the network with only excitatory couplings (Fig.~\ref{Fig5}(c),(d)). 
What follows that a further increase in $p_{\mathrm {in}}$ leads to an 
increase in $\overline{C}$ for a
complete inhibitory networks, which can be explained from the
fact that for very high $p_{\mathrm in}$ values, it is very difficult for
the evolved networks to attain balance of inhibitory and excitatory connection 
weight, thus leading to the evolution of $\overline{C}$ to higher values.
This discussion will become more clear
in the following section where in order to provide an insight
to the emergence of clustering we provide a detailed comparison of $R_{max}$
values of regular lattices with the corresponding random networks. While
making this comparison, it is appropriate to consider random
fluctuations in connection strength of the regular lattices
in the same line as done for the random network model considered here.

Fig.~\ref{Fig5} presents spread of $R_{max}$ about its mean over many
realizations. Upon comparing networks having only excitatory coupling, the mean
value of $R_{max}$ for lattices (Fig.~\ref{Fig5}(a)) are lesser than
those of the random networks (Fig.~\ref{Fig5}(b)). 
What follows that if the overlapping region between the ranges of $R_{\mathrm {max}}$ values in lattices and random networks is broader, evolution of clustering is hampered.
In case of systems with only inhibitory couplings (Fig.~\ref{Fig5}(e),(f)), the overlapping region is less, as a result of which significant clustering behavior is observed, as opposed to the cases with only excitatory 
connections (Fig.~\ref{Fig5}(a) and (b)).
For the mixed couplings case ($p_{\mathrm {in}}$ = 0.50), 
the above explanation does not stand valid as here the 
clustering evolves in spite of the overlapping region being broader (Fig. 4(c)
and 4(d)). This behavior might be driven by the fact that $\overline{C}_{\mathrm {corr}}$ for the mixed case
adopts higher negative values in the evolved networks as compared
to the excitatory and inhibitory networks.
\begin{figure}[h]
\centerline{\includegraphics[height=3.5cm,width=4cm]{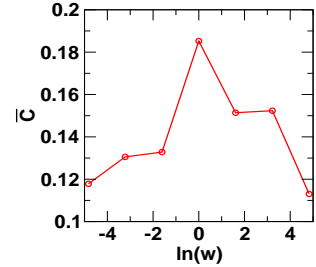}}
\caption{(Color online) Impact of $w$ on the evolved value of $\overline{C}$.
 Systems are evolved upto 5000 generations and average value
is taken over last 1000 generations of population networks used in the GA.
For each case average degree of network is 6, number of node is 100 and  $p_{\mathrm{in}}$ = 0.50.}
\label{Fig6}
\end{figure} 
Further, inhibitory weight ($w$)  in [$c_{\mathrm {ij}}$] matrices 
can be introduced as follows;
\begin{equation}
d_{\mathrm {ij}} = \begin{cases} c_{\mathrm {ij}}w~~\mbox{if } c_{\mathrm {ij}}<0 \\
c_{\mathrm {ij}}. \end{cases}
\end{equation}
For $w<1$, the networks have higher excitatory coupling strength as 
compared to the inhibitory coupling strength, whereas for  $w>1$,  the
vice-versa is true. We vary the value of $w$ while keeping
inhibitory-excitatory coupling ratio fixed to $1:1$,
as for the higher or lower values of inhibition,
the networks evolve towards the balanced condition (Fig.~\ref{Fig2}b).
As $w$ increases, the inhibitory coupling strength
increases and becomes equal to the excitatory strength at $w = 1$.
A further increase in $w$ leads to the 
higher inhibitory coupling strength as compared to the
excitatory and in turn, inclusion of $w$ affects properties of evolved
matrices as well. With an increase in $w$,  
$\overline{{C}}$ increases until $w = 1$. Further increase in the value
of $w$ 
leads to a decrease in the $\overline{C}$ values (Fig.~\ref{Fig6}).
Therefore, the balanced coupling strength yields the
highest clustering values in the evolved networks.
 
We remark that
the Watts-Strogatz model was proposed to capture high clustering and
low diameter properties possessed by many real world networks \cite{Strogatz}.
This model, based on rearrangement of connections among nodes, leads to
the networks having high clustering values.
In the similar lines, using simulated annealing technique clustering
have been shown to be evolved as a consequence of tradeoff between maximized connectivity
and minimized wiring \cite{Mathias_PRE_2001}.
Further, triadic closure mechanism
which is based on enforcement of the connections to form triads models
clustering coefficient distribution of the triads over the nodes observed in
many real networks \cite{Alexi}. Various other models
motivated from the closer triadic method produce
other realistic interaction patterns \cite{Ginestra, Davidsen, Marsili, Jackson}. All these
methods take one or other parameter to create the abundance of triads, whereas
our work uses a very different approach. In this approach the
initial random networks are evolved to a structure 
having realistic features by selecting those networks during the evolution
which have structural properties beneficial for a stable dynamics. 
Additionally, this approach 
does not enforce the connections to form
triads \cite{Alexi}, or does not start with a clustered network \cite{Strogatz},
infact the clustering evolves naturally in a network.
Another advantage of the method 
is to provide an explanation to the abundance of omnivory-chain
existing in nature \cite{Stouffer,Jordi}.

To conclude, we present a novel method based on GA to evolve a network
which has high clustering. 
We demonstrate that as a system proceeds towards stability maximization
accounted by $R_{max}$, clustering coefficient also follows an
increasing trend implicating its importance for the stability of the
system during its evolution. 
Presence of inhibitory links, among other parameters such
as fluctuations in coupling
strength and the predefined interaction pattern between given pair of
nodes, emerges as crucial factors in evolution of clustering.
In the case of ER network, the expected values of $C$ are equal to 
its connection probability. However, real world networks are known to have
high degree of clustering due to their non random local structures. Randomness and optimization 
features co-exist in complex real world systems \cite{barabasiluck_reason}.
We present the evolution of clustering in random networks by maximizing the stability using an optimization technique, which is close to GA.
Connectivity is a constraint that a system ought to maintain in order to attain completeness, and need not be a mandatory feature from the stability view point.
A connection behavior is dependent on its interacting individuals, the 
nature of which, as suggested in this paper, will remain predefined in case any future connection happens to arise between them. 
Further, the coupling strengths might carry  fluctuations as randomness is a universal phenomenon \cite{MayNature1972,Allesina,Sompolinsky}. 
With this assumptions, the simulation results provide plausible reasons behind the origin of clustering and clique distribution prevalent in real world systems. 
These features further help in attaining hierarchical organization of modularity with clustering behavior. They also help in unraveling the evolutionary rules behind the existence of local motif structures having cliques of order three.
While importance of inhibition has already been emphasized for functioning
and evolution (see for example \cite{Brain,Extreme}), this 
paper demonstrates that inhibition is crucial for the evolution of clustering, with
an additional essential parameter which is randomness. Even in the
case of systems with only inhibitory couplings, clustering exists in absence of randomness in coupling strength. Randomness in coupling
 weight yields a strong clique formation when inhibition is present whereas for only excitatory couplings, the effect is weak.  As long as initial 
networks have very poor clustering, which can be treated as an adverse situation, there is a scope of evolution through GA by maximizing 
stability which further detects the importance of inhibition. For instance, if we start with random networks having high average degree 
which implicates in clustering as high as comparable to real world networks.

The main assumption behind the formation of the behavior matrix
[$b_{\mathrm{ij}}$] is the time invariant behavior of coupling. However
small changes in nature of interaction in behavior matrix, with some
minimal probability which propagates during the evolution, leads
to clustering, provided the matrix obeys the main constraint of the presence of inhibitory coupling.
This establishes a flexibility of the model for explaining the origin of high clustering value
evident in various systems. With an increase in the probability of introducing 
changes in the behavior, we have observed that $R_{\mathrm max}$
values cease to decrease with no significant increase in the $C$ values 
clearly indicating the
importance of a fixed coupling behavior in the course
of evolution providing an interesting segment to be explored in future. 
We remark that while increase in system size and number of connections
($N$ and $\langle k \rangle$) remain an important aspect  
\cite{Alexi}, 
change in the interaction pattern or rewiring is also considered to
be crucial for evolution \cite{Vallender_2008}. The scheme presented in this paper considered fixed values of $N$ and number of connections throughout the 
evolution in order to capture impact of the
stability maximization on the behavior of
interaction pattern.

Clustering is a feature common in both man-made and natural complex systems. In man-made systems, several factors influence the emergence of clustering, for example in social networks, making friendship with people having common
friends due to their common professions or common opinions, is highly probable. Such a provision of commonness does not seem to exist in natural systems. Stability can be treated as a factor while modeling both man-made and natural systems.
Though the importance of inhibition in the evolution of clustering through stability maximization has been explained using
very simple model system of 1-d lattice, the results
presented here first time reveal the importance of inhibitory coupling for evolution of
clustering using GA.
This framework thus opens a new interesting direction
towards understanding the evolution of structures in complex networks where 
inhibition in connections is not only present but has been found to be 
crucial for functioning 
of underlying system, for
instance brain networks and ecological networks. This optimization technique,
inspired by the Darwinian evolution,
can be further extended to 
more general setting in order to get insight into evolutionary origin of other structural properties \cite{bipartite}.
Our approach, although not exactly but up to a satisfiable extent could evolve the omnivory chain prevalent in ecological systems \cite{Stouffer,Jordi}. Optimization of the stability maximization approach in order to replicate the real world scenario would certainly be an interesting future direction to investigate. 

SJ is grateful to DST  
(SR/FTP/PS-067/2011) and CSIR (25(02205)/12/EMR-II) for financial support.
SKD acknowledges the University grants commission, India for
financial support and members of complex systems lab for timely help and
useful discussions.

\end{document}